\documentclass[amsmath,amsfonts,amssymb,twocolumn,nofootinbib]{revtex4-1}
\usepackage{graphicx}
\usepackage{verbatim}
\usepackage{hyperref}
\usepackage[table]{xcolor}
\usepackage{algcompatible}
\usepackage[floatrow]{trivfloat}

\trivfloat{algorithm}

\newcommand{\sabs}[1]{\ensuremath { \raisebox{1pt}{$\wr$} \mathbf{#1} \raisebox{1pt}{$\wr$} } }
\newcommand{\sabsGreek}[1]{\ensuremath { \raisebox{1pt}{$\wr$} \mbox{\boldmath{$#1$}} \raisebox{1pt}{$\wr$} } }

\newcommand{\tabHead}[1]{  \begin{minipage}[t]{2cm} \textbf{#1} \end{minipage} }

\definecolor{comment}{gray}{0.50}

\begin{document}

\title{A General Metric for Riemannian Manifold Hamiltonian Monte Carlo}
\author{Michael Betancourt}
\affiliation{Department of Statistics, Columbia University, New York, NY 10027, USA}
\email{betanalpha@gmail.com}

\date{\today}

\begin{abstract}

Markov Chain Monte Carlo (MCMC) is an invaluable means of inference with complicated models, and Hamiltonian Monte Carlo, in particular Riemannian Manifold Hamiltonian Monte Carlo (RMHMC), has demonstrated impressive success in many challenging problems.  Current RMHMC implementations, however, rely on a Riemannian metric that limits their application to analytically-convenient models.  In this paper I propose a new metric for RMHMC without these limitations and verify its success on a distribution that emulates many hierarchical and latent models.

\end{abstract}


\maketitle

Riemannian Manifold Hamiltonian Monte Carlo provides a powerful tool for the efficient sampling from complex distributions, but the applicability of existing approaches has been limited by the dependency on the Fisher-Rao metric.  In this paper I introduce a new metric that admits a general implementation of Riemannian Manifold Hamiltonian Monte Carlo and demonstrate its efficacy on a distribution that mirrors the pathological behavior of common models.

\section{Hamiltonian Monte Carlo}

Hamiltonian Monte Carlo (HMC) takes advantage of symplectic geometry to yield efficient Markov transitions~\cite{Betan2011}.  Augmenting the parameters of an $N$-dimensional target density, $\pi \! \left( \mathbf{q} \right)$, with corresponding momenta, $\mathbf{p}$, defines a joint density,
\begin{align*}
\pi \! \left( \mathbf{p}, \mathbf{q} \right) &= \pi \! \left( \mathbf{p} | \mathbf{q} \right) \pi \! \left( \mathbf{q} \right) \\
&= \exp \left[ \log \pi \! \left( \mathbf{p} | \mathbf{q} \right) \right] \exp \left[ \log \pi \! \left( \mathbf{q} \right) \right] \\
& \propto \exp \left[ - T \! \left( \mathbf{p}, \mathbf{q} \right) \right] \exp \left[ - V \! \left( \mathbf{q} \right) \right] \\
&\equiv \exp \left[ - H \! \left( \mathbf{p}, \mathbf{q} \right) \right].
\end{align*}

The Hamiltonian, $H \! \left( \mathbf{p}, \mathbf{q} \right) = T \! \left( \mathbf{p}, \mathbf{q} \right) + V \! \left( \mathbf{q} \right)$, defines trajectories between points $\mathbf{z} = \left\{ \mathbf{p}, \mathbf{q} \right\}$ via the differential equations
\begin{align*}
\frac{d \mathbf{q} }{dt} &= + \frac{ \partial H }{ \partial \mathbf{p} } \\
\frac{d \mathbf{p} }{dt} &= - \frac{ \partial H }{ \partial \mathbf{q} }.
\end{align*}
Because these trajectories preserve the value of the Hamiltonian and the differential volume $\mathrm{d}^{2N} \mathbf{z}$, they also define Markovian transitions with the stationary density $\pi \! \left( \mathbf{p}, \mathbf{q} \right)$.  Alternating this Hamiltonian evolution with conditional samples of the momenta,
\begin{equation*}
\mathbf{p} \sim \pi \! \left( \mathbf{p} | \mathbf{q} \right) \propto \exp \left[ - T \! \left( \mathbf{p}, \mathbf{q} \right) \right],
\end{equation*}
yields an ergodic Markov chain sampling from $\mathbf{z}$ and, because the marginal of $\pi \! \left( \mathbf{p}, \mathbf{q} \right)$ is constructed to be the target distribution, the desired samples from $\pi \! \left( \mathbf{q} \right)$ follow by simply disregarding the momenta.

No matter the choice of the kinetic energy, $T \! \left( \mathbf{p}, \mathbf{q} \right)$, the evolution equations incorporate the gradient of the potential, $V \! \left( \mathbf{q} \right)$, and hence higher order information about the target distribution.  This gradient guides the Markov chain along regions of high probability mass and reduces random walk behavior.  Note that, in practice, the Hamiltonian evolution cannot be performed analytically and we must resort to numerical integration.  Error in the integration scheme introduces bias into the transitions, but this is readily avoided by considering the evolution not as a transition but rather as the proposal for a Metropolis transition~\cite{Duane1987, Neal2011}.

The first~\cite{Duane1987} and still most common choice of the conditional density, $\pi \! \left( \mathbf{p} | \mathbf{q} \right)$, is a standard gaussian,
\begin{equation*}
\pi \! \left( \mathbf{p} | \mathbf{q} \right) = \mathcal{N} \! \left( \mathbf{p} | \mathbf{0}, \mathbf{M} \right),
\end{equation*}
or
\begin{equation} \label{EMHMC}
 T \! \left( \mathbf{p}, \mathbf{q} \right) = \frac{1}{2} \mathbf{p}^{T} \cdot \mathbf{M}^{-1} \cdot \mathbf{p},
 \end{equation}
where the mass matrix $\mathbf{M}$ allows for a global decorrelation and rescaling of the parameters with respect to each other.  This choice, however, ultimately limits the effectiveness of HMC when applied to intricate target distributions.  Because $ \mathbf{p}^{T} \cdot \mathbf{M}^{-1} \cdot \mathbf{p}$ is a $\chi^{2}$ variate, in equilibrium $\Delta T \approx N / 2$ and, with the Hamiltonian conserved along each trajectory, this implies that the variation in the potential is also limited to $\Delta V \approx N / 2$.  When the target distribution is highly correlated, the typical set spans a potential gap much larger than this: the resulting samples become highly correlated no matter how long the trajectories are evolved~\cite{Neal2011} and the Markov chain devolves towards a random walk.  

Another issue with the simple choice \eqref{EMHMC} is that the inevitable numerical integration introduces a spatial scale into the system via a finite step-size.  Complicated target distributions will typically exhibit multiple spatial scales depending on the particular value of the parameters, and any single choice of a step-size will generate at least some inefficiency.  If the step-size is chosen to maximize efficiency, as common in adaptive schemes, regions of the target distribution with large curvature, and hence small spatial scales, can be missed entirely by the numerical trajectories.

These weaknesses can be overcome by appealing to a more sophisticated choice of the conditional density: a gaussian conditionally dependent on the $\mathbf{q}$ through a covariance matrix,
\begin{equation*}
\pi \! \left( \mathbf{p} | \mathbf{q} \right) = \mathcal{N} \! \left( \mathbf{p} | \mathbf{0}, \mathbf{\Sigma} \! \left( \mathbf{q} \right) \right),
\end{equation*}
or
\begin{equation*}
T \left( \mathbf{p}, \mathbf{q} \right) = \frac{1}{2} \mathbf{p}^{T} \cdot \mathbf{\Sigma}^{-1} \! \left( \mathbf{q} \right) \cdot \mathbf{p} + \frac{1}{2} \log | \mathbf{\Sigma} \! \left( \mathbf{q} \right) |.
\end{equation*}
Because the resulting Hamiltonian trajectories are related to geodesics on a Riemannian manifold with metric $\mathbf{\Sigma} \! \left( \mathbf{q} \right)$, this choice is known as \textit{Riemannian Manifold Hamiltonian Monte Carlo} (RMHMC)~\cite{Girolami2011}.  Similarly, the constant metric of \eqref{EMHMC} can be thought of as emulating dynamics on a Euclidean manifold, and to be consistent I will refer to use of the simpler Hamiltonian as \textit{Euclidean Manifold Hamiltonian Monte Carlo} (EMHMC).

The freedom in specifying a metric admits two significant improvements: a proper choice of $\mathbf{\Sigma} \! \left( \mathbf{q} \right)$ can dynamically decorrelate and rescale the target distribution to avoid inefficiencies in the numerical integration, while also yielding a dynamic determinant whose variations can compensate for much larger variations in the potential.

What, however, exactly defines a proper choice for the metric?  When the target distribution is a multivariate gaussian,
\begin{equation*}
V \! \left( \mathbf{q} \right) = \frac{1}{2} \mathbf{q}^{T} \cdot \mathbf{S}^{-1} \cdot \mathbf{q},
\end{equation*}
the target distribution is standarized by taking $\mathbf{\Sigma} \! \left( \mathbf{q} \right) = \mathbf{S}^{-1}$~\cite{Neal2011}.  In a convex neighborhood any target distribution can be approximated by a multivariate gaussian,
\begin{equation*}
\pi \! \left( \mathbf{q} \right) \approx \mathcal{N} \! \left( \mathbf{q} | \mathbf{0}, \mathbf{H}^{-1} \right)
\end{equation*}
or, equivalently,
\begin{equation*}
V \! \left( \mathbf{q} \right) \approx \frac{1}{2} \mathbf{q}^{T} \cdot \mathbf{H} \cdot \mathbf{q},
\end{equation*}
with the Hessian matrix
\begin{equation*}
H_{ij} = \partial^{2} V / \partial q^{i} \partial q^{j},
\end{equation*}
which immediately motivates the candidate metric $\mathbf{\Sigma} \! \left( \mathbf{q} \right) = \mathbf{H}$.  

This metric quickly runs into problems, however, when the target distribution is not globally convex.  In neighborhoods where the Hessian is not positive-definite, for example, the conditional density $\pi \! \left( \mathbf{p} | \mathbf{q} \right)$ becomes improper.  Moreover, in the neighborhoods where the signature of the Hessian changes, the log determinant diverges and the Hamiltonian evolution becomes singular.  These neighborhoods effectively partition the support of the target distribution into a disjoint union of compact neighborhoods between which the Markov chain cannot transition.

One way to avoid indefinite metrics is to take advantage of any conditioning variables, $\mathbf{y}$, in the target distribution.  Marginalizing the Hessian over these conditioning variables yields the Fisher-Rao metric~\cite{Amari2007},
\begin{equation*}
\Sigma_{ij} = \mathbb{E}_{\mathbf{y}} \left[ \frac{ \partial^{2} V \! \left( \mathbf{q} | \mathbf{y} \right) }{ \partial q^{i} \partial q^{j} } \right],
\end{equation*}
which is guaranteed to be positive-semidefinite.  For all but the simplest conditional distributions, however, the marginalization is unfeasible and, even when it can be performed analytically, the resulting metric can still be singular.  Moreover, the marginalization removes the correlation between variables in many hierarchical and latent models, almost eliminating the effectiveness of the metric.  Of course, all of this is immaterial if the target distribution lacks natural conditioning variables.

We need a means of constructing a metric from the Hessian that is not only everywhere well-behaved but also practical to compute for any given target distribution.

\section{The SoftAbs Metric}

With a careful application of matrix functions, it is possible to maintain the desirable behavior of the Hessian in convex neighborhoods while avoiding its singular behavior elsewhere.  Moreover, because the functions are local the resulting metric is readily implemented for general distributions.

\subsection{Definition}

The exponential map~\cite{Spivak2005a}, $\exp$, is a matrix function from the space of all matrices to the component of the general linear group, $\mathrm{GL} \! \left(n\right)$, connected to the identity matrix: an isomorphism of the space of positive-definite matrices.  Because this mapping preserves the symmetric part of the domain, any symmetric matrix, such as the Hessian, is guaranteed to be mapped to a symmetric, positive-definite matrix admissible as a Riemannian metric.

One benefit of the exponential map is that it preserves the eigenbasis of the input matrix, $\mathbf{X}$.  If 
\begin{equation*}
\mathbf{X} = \mathbf{Q} \cdot \mbox{\boldmath{$\lambda$}} \cdot \mathbf{Q}^{T}
\end{equation*}
is the eigendecomposition of $\mathbf{X}$ with $\mbox{\boldmath{$\lambda$}} = \mathrm{Diag} \left( \lambda_{i} \right)$ the diagonal matrix of eigenvalues and $\mathbf{Q}$ the corresponding matrix of eigenvectors, then the exponential map yields
\begin{equation*}
\exp \! \mathbf{X} = \mathbf{Q} \cdot \exp \! \mbox{\boldmath{$\lambda$}} \cdot \mathbf{Q}^{T}.
\end{equation*}
The metric $\exp \! \mathbf{H}$ provides the same decorrelation as the Hessian but, unfortunately, also severely warps the eigenvalues and the corresponding rescaling of the local parameters.  

\begin{figure}
\centering
\includegraphics[width=3in]{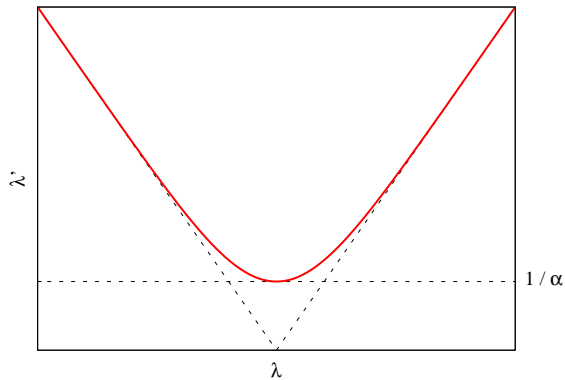}
\caption{The SoftAbs map preserves the eigenbasis of the Hessian but transforms the eigenvalues, $\lambda$, with a smooth approximation to the absolute value.  The inverse of the regularization parameter, $\alpha$, controls the ``hardness'' of the approximation; as $\alpha \rightarrow \infty$ the SoftAbs map reduces to the exact absolute value.
\label{fig:softAbsEigen}}
\end{figure}

By combining multiple exponential mappings, however, we can largely preserve the spectral decomposition of the Hessian.  In particular, the \textit{SoftAbs} map
\begin{align*}
\sabs{X} & \equiv \\
&\left[ \exp \! \left( \alpha \mathbf{X} \right) + \exp \! \left( -\alpha \mathbf{X} \right) \right] \cdot \mathbf{X} \cdot \left[ \exp \! \left( \alpha \mathbf{X} \right) - \exp \! \left( - \alpha \mathbf{X} \right) \right]^{-1}
\end{align*}
approximates the absolute value of the eigenspectrum with a smooth function:
\begin{equation*}
\sabs{X} = \mathbf{Q} \cdot \sabsGreek{\lambda} \cdot \mathbf{Q}^{T},
\end{equation*}
where
\begin{align*}
\sabsGreek{\lambda} &= \mathrm{Diag} \left( \lambda_{i} \frac{ e^{ \alpha \lambda_{i} } + e^{ -\alpha \lambda_{i} }  }{ e^{ \alpha \lambda_{i} } - e^{ - \alpha \lambda_{i} } }  \right) \\
&= \mathrm{Diag} \left( \lambda_{i} \coth \alpha \lambda_{i} \right).
\end{align*}
This map not only ensures that the transformed eigenvalues are positive but also regularizes any small eigenvalues that might introduce numerical instabilities (Figure \ref{fig:softAbsEigen}).  

Applying the SoftAbs map to the Hessian guarantees a well-behaved metric for RMHMC, $\sabs{H}$, that preserves the desired properties of the Hessian while regularizing its numerical singularities.  In a practical implementation, $\alpha$ limits the scaling of the integration step-size and restrains the integrator from unwise extrapolations, emulating a trust region common in nonlinear optimization~\cite{Celis1985}.

This construction also motivates a family of approximate metrics with possible utility in circumstances of limited computational resources (Appendix~\hyperref[sec:approximations]{A}).

\subsection{Implementation}

In practice, exponential maps can be difficult to implement~\cite{Moler2003}; the eigendecomposition used above, for example, can suffer from numerical instabilities when applied to general matrices because of ambiguities among the eigenvectors.  The Hessian, however, is symmetric and the eigenvectors are guaranteed to be orthogonal.  Consequently, the eigendecomposition is well-behaved and provides a practical means of computing the SoftAbs map.  

To implement the SoftAbs metric we first perform the eigendecomposition of the Hessian
\begin{equation*}
\mathbf{H} = \mathbf{Q} \cdot \mbox{\boldmath{$\lambda$}} \cdot \mathbf{Q}^{T},
\end{equation*}
and then reconstruct the metric as
\begin{equation*}
\sabs{H} = \mathbf{Q} \cdot \sabsGreek{\lambda} \cdot \mathbf{Q}^{T},
\end{equation*}
with $\sabsGreek{\lambda} = \mathrm{Diag} \left( \lambda_{i} \coth \alpha \lambda_{i} \right)$.

Hamiltonian evolution also requires two derivatives: the gradient of the quadratic form, $\mathbf{p}^{T} \cdot \sabs{H}^{-1} \cdot \mathbf{p}$, and the log determinant, $\log \left| \sabs{H} \right|$.

The latter can be computed as~\cite{Aizu1963, Wilcox1967}
\begin{align*}
\partial \left( \mathbf{p}^{T} \cdot \sabs{H}^{-1} \cdot \mathbf{p} \right) &= - \mathbf{p}^{T} \cdot \sabs{H}^{-1} \cdot \partial \sabs{H} \cdot \sabs{H}^{-1} \cdot \mathbf{p} \\
&= - \mathbf{M}^{T} \left[ \mathbf{J} \circ \mathbf{Q}^{T} \cdot \partial \mathbf{H} \cdot \mathbf{Q} \right] \mathbf{M},
\end{align*}
where $\circ$ denotes the Hadamard product,
\begin{equation*}
\mathbf{M} = \sabsGreek{\lambda}^{-1} \cdot \mathbf{Q}^{T} \cdot \mathbf{p},
\end{equation*}
and
\begin{equation*}
J_{ij} \equiv \frac{ \lambda_{i} \coth \alpha \lambda_{i} - \lambda_{j} \coth \alpha \lambda_{j} }{ \lambda_{i} - \lambda_{j} }.
\end{equation*}
Note that when $\lambda_{i} = \lambda_{j}$, such as for the diagonal elements or degenerate eigenvalues, this becomes the derivative,
\begin{equation*}
J_{ij} \rightarrow \frac{ \partial }{ \partial \lambda_{i} } \lambda_{i} \coth \alpha \lambda_{i}.
\end{equation*}

Unfortunately, this form of the gradient is computationally inefficient, requiring $O \! \left( N^{3} \right)$ for each component of the gradient, and hence $O \! \left( N^{4} \right)$ overall.  Taking advantage of the properties of the Hadamard product~\cite{Magnus2007}, however, the gradient can be manipulated to give
\begin{align*}
\partial \left( \mathbf{p}^{T} \sabs{H}^{-1} \mathbf{p} \right) &= - \mathrm{Tr} \left[ \mathbf{Q} \cdot \mathbf{D} \cdot \mathbf{J} \cdot \mathbf{D} \cdot \mathbf{Q}^{T} \cdot \partial \mathbf{H} \right],
\end{align*}
where $\mathbf{D} = \mathrm{Diag} \left( \left( \mathbf{Q}^{T} \cdot \mathbf{p} \right)_{i} / \lambda_{i} \coth \alpha \lambda_{i}\right)$.  If the matrix $\mathbf{Q} \cdot \mathbf{D} \cdot \mathbf{J} \cdot \mathbf{D} \cdot \mathbf{Q}^{T}$ is first cached, then each component of the gradient can be computed in only $O \! \left( N^{2} \right)$ so that the complete gradient does not exceed the $O \! \left( N^{3} \right)$ complexity of the decomposition itself.

Similar Hadamard identities reduce the gradient of the log determinant to
\begin{align*}
\partial \log \left | \wr \mathbf{H} \wr \right| &= \mathrm{Tr} \left[ \mathbf{Q} \left( \mathbf{R} \circ \mathbf{J} \right) \mathbf{Q}^{T} \cdot \partial \mathbf{H} \right],
\end{align*}
where
\begin{equation*}
\mathbf{R} = \mathrm{Diag} \left( \frac{1}{ \lambda_{i} \coth \alpha \lambda_{i} } \right).
\end{equation*}
Once again, caching the intermediate matrix, $\mathbf{Q} \left( \mathbf{R} \circ \mathbf{J} \right) \mathbf{Q}^{T}$, enables the full gradient to be computed in $O \! \left( N^{3} \right)$.

These results admit an efficient symplectic integrator (Appendix~\hyperref[sec:integrator]{B}) for RMHMC; a C++ implementation is available online at \url{http://betanalpha.github.com/jamon/}.

\begin{table*}[t!]
\begin{center}
	\renewcommand{\arraystretch}{1.2}
	\begin{tabular}{c@{}c@{}c@{}c@{}c@{}c@{}c@{}c}
	\rowcolor[gray]{0.8} \tabHead{Algorithm} & \tabHead{Warm-Up Iterations} & \tabHead{Samples} & \tabHead{$\mathbf{\epsilon}$} & \tabHead{Accept Rate} & \tabHead{CPU Time (s)} & \tabHead{ESS} & \tabHead{ESS / Time ($\mathbf{s^{1}}$)} \\
	 EMHMC & $10^{3}$ & $10^{5}$ & 0.001 & 0.999 & 1627 & 70.3 & 0.0432 \\
	 \rowcolor[gray]{0.8} RMHMC & $10^{3}$ & $10^{3}$ & 0.21 & 0.946 & 6282 & 856 & 0.136 \\
	\end{tabular}
	\caption{When comparing the effective sample size of the latent variable, $v$, in the funnel distribution, hand-tuned EMHMC is over three times less effective than adaptively-tuned RMHMC.  CPU time was measured with the \texttt{clock} function in the C++ library \texttt{time}.}
	\label{tab:benchmark}
\end{center}
\end{table*}

\section{Experiments}

The utility of the SoftAbs metric is best demonstrated on complex distributions.  Neal's funnel distribution~\cite{Neal2003}
\begin{equation*}
\pi \! \left( \mathbf{x}, v \right) = \prod_{i = 1}^{n} \mathcal{N} \! \left( x_{i} | 0, e^{-v} \right) \cdot \mathcal{N} \! \left( v | 0, 9 \right),
\end{equation*}
emulates many pathological features of popular distributions, such as those arising in hierarchical~\cite{Gelman2004} and latent~\cite{Murray2010} models.  Note that, by construction, the marginal distribution of  $v$ is simply $v \sim \mathcal{N} \! \left(0, 9 \right)$\footnote{Note the use of the convention $\mathcal{N} \! \left( \mu, \sigma^{2} \right)$.}, independent of $n$,  admitting $v$ and its marginal distribution as a simple diagnostic of bias in any sampling procedure.

In each experiment a Markov chain is randomly initialized, $q_{i} \sim U \! \left(-1, 1\right)$, and then taken through a series of warm-up iterations before sampling begins.  Where noted, the integrator step-size, $\epsilon$, is adapted with dual-averaging to yield a target Metropolis acceptance rate~\cite{Hoffman2011}.  The number of integration steps is set by hand to approximate the half-period of the oscillating trajectories (Figures \ref{fig:noadaptFlatFunnelDiptych}, \ref{fig:softAbsFunnelDiptych}).

Autocorrelations, $\rho_{i}$, of $v$ are computed with an initial monotone sequence estimator~\cite{Geyer2011} and the effective sample size (ESS) is defined as $\mathrm{ESS} = I \left( 1 + 2 \sum_{i = 1}^{I} \rho_{i} \right)^{-1}$, where $I$ is the total number of generated samples.

The above procedure is applied to EMHMC with step-size adaptation, EMHMC without step-size adaption, and RMHMC with the SoftAbs metric.

\subsection{EMHMC with Adaptation}

\begin{figure}[h!]
\centering
\includegraphics[width=7in]{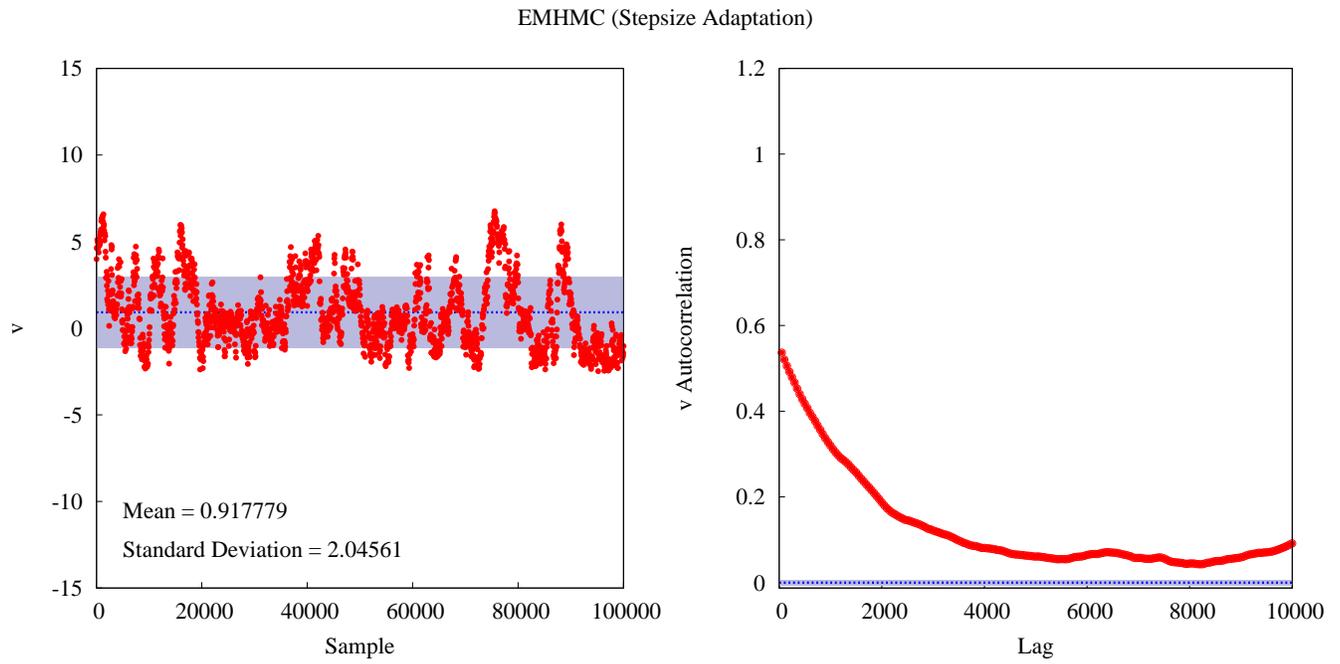}
\caption{Adaptive tuning of EMHMC results in an excessively large step-size, preventing trajectories from exploring the neck of the funnel and biasing the stationary distribution.  Consequently, the samples of $v$ are inconsistent with the marginal distribution $\mathcal{N} \! \left( 0, 9 \right)$.
\label{fig:adaptFlatFunnelTrace}}
\end{figure}

Despite its simplicity, the funnel demonstrates many of the limitations of EMHMC.  When adaptively tuned to the nominal acceptance rate $r = 0.65$~\cite{Neal2011}, the integrator step-size exceeds the spatial scale of the narrow neck; even though the probability mass of the mouth and neck of the funnel is comparable, the resulting trajectories overlook the neck entirely and bias resulting expectations without any obvious indication (Figure \ref{fig:adaptFlatFunnelTrace}).  

\subsection{EMHMC without Adaptation}

\begin{figure}[h!]
\centering
\includegraphics[width=7in]{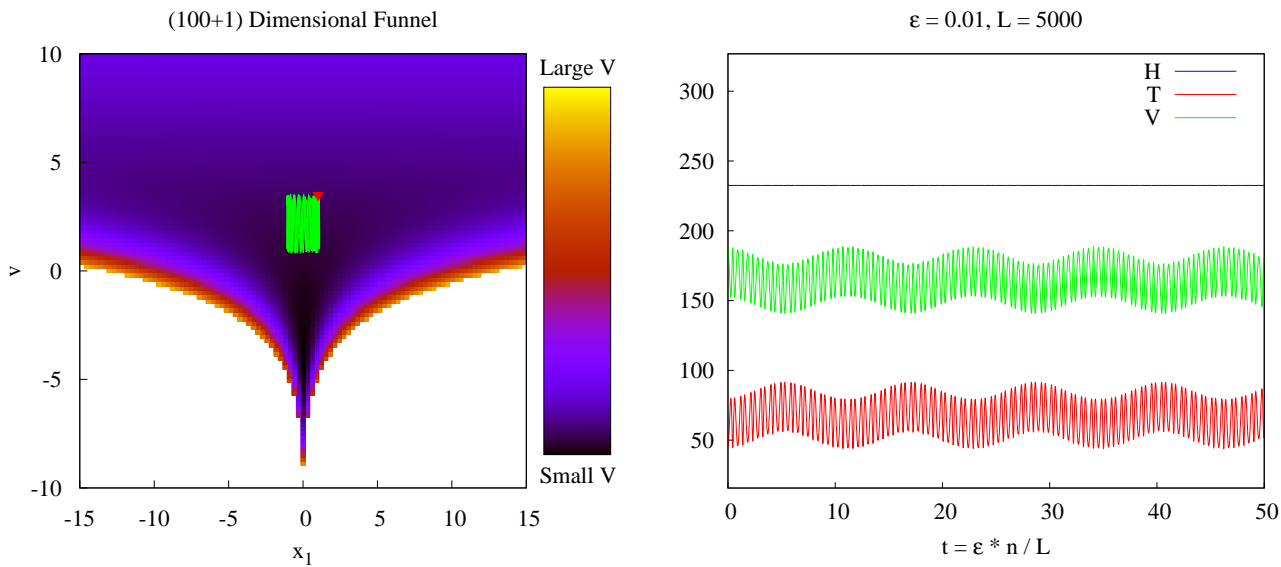}
\caption{EMHMC trajectories are limited to $\Delta V \sim (n + 1) / 2$ and consequently explore only a small neighborhood of the funnel.  Note that the trajectories are oscillatory; half of the largest period of oscillation, here $T / 2 \approx 8$, defines the optimal integration time for maximizing distance, and minimizing autocorrelation, between the samples.
\label{fig:noadaptFlatFunnelDiptych}}
\end{figure}

\begin{figure}[h!]
\centering
\includegraphics[width=7in]{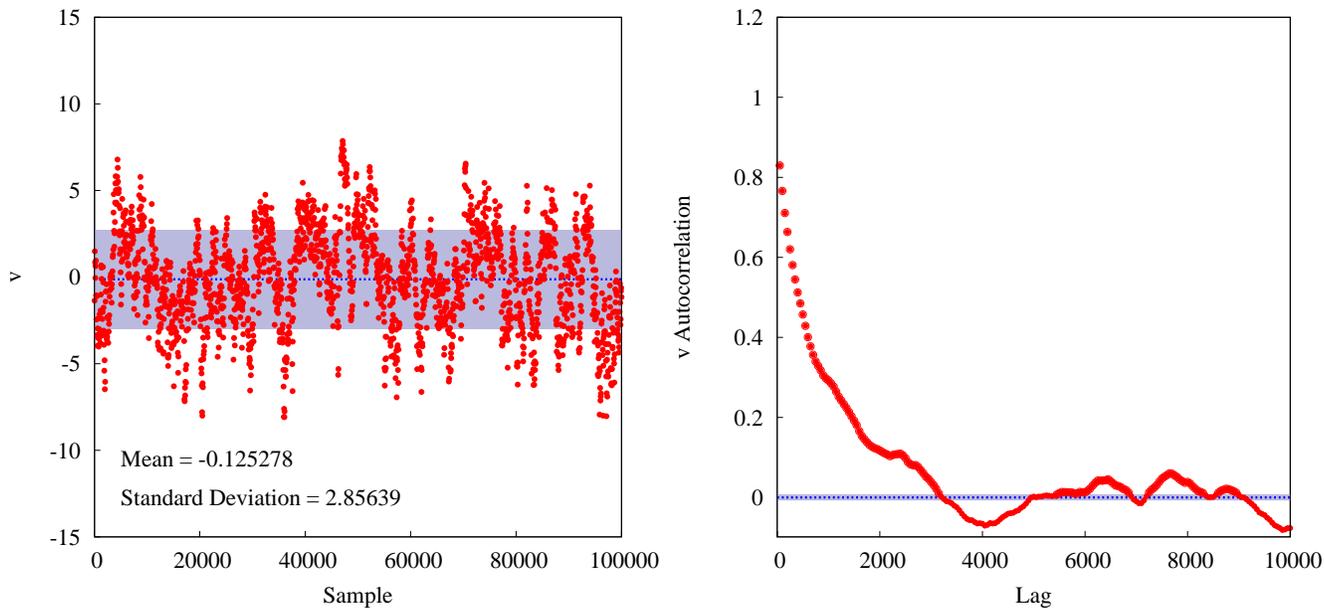}
\caption{Although not optimal, a smaller step-size tuned by hand allows the trajectories to penetrate the neck of the funnel, and the resulting samples of $v$ are consistent with the true marginal, $\mathcal{N} \! \left(0, 9 \right)$.  In real applications where the truth is not known a priori, this sort of hand-tuning is not viable.
\label{fig:noadaptFlatFunnelTrace}}
\end{figure}

Because we know the truth in this case, we can abandon adaptive tuning and instead tune the step-size by hand; a smaller step-size ensures that the trajectories explore most of the funnel's probability mass and that the marginal distribution $p \! \left( v \right)$ is correct within Monte Carlo error.  Unfortunately, the funnel also exhibits the limitations of a position-independent kinetic energy.  The variation of the potential within the typical set is huge, and the meager variation of the kinetic energy dramatically restricts the distance of each transition (Figure \ref{fig:noadaptFlatFunnelDiptych}).  The EMHMC transitions struggle to cross between the mouth and neck of the funnel, and the Markov chain becomes little more than a random walk across the distribution (Figure \ref{fig:noadaptFlatFunnelTrace}).  

\subsection{RMHMC with the SoftAbs Metric}

\begin{figure}[h!]
\centering
\includegraphics[width=7in]{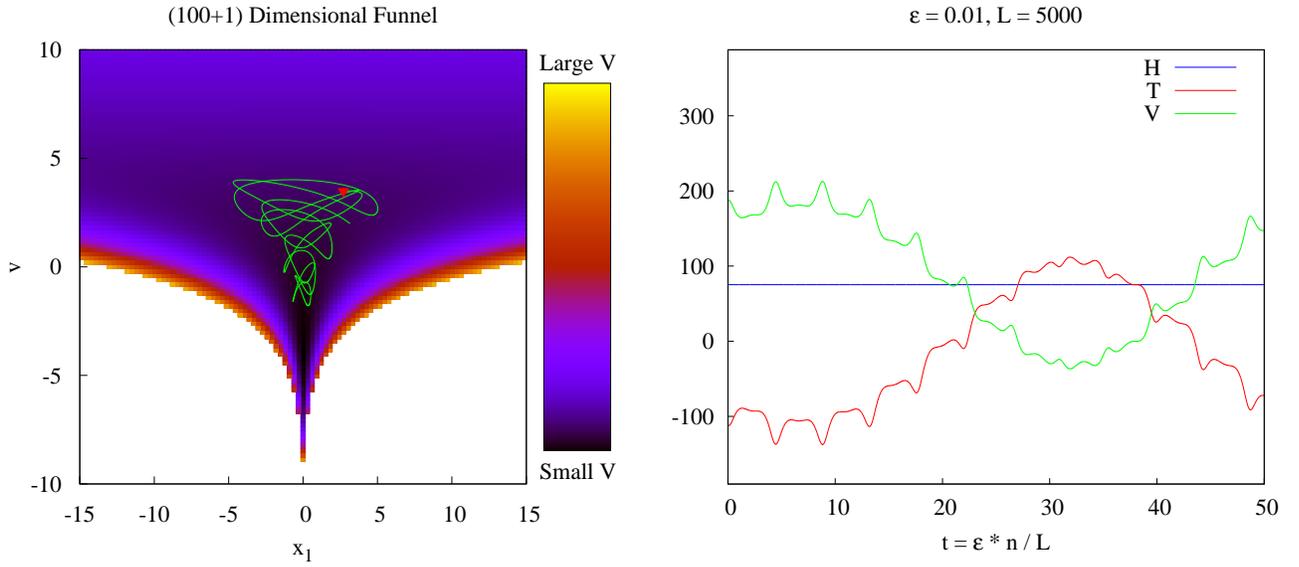}
\caption{The log determinant in the Hamiltonian allows RMHMC trajectories to explore without the restriction to ${\Delta V \sim (n +1) / 2}$, and the dynamic decorrelation/scaling ensures that a single integrator step-size is efficient across the entire distribution.  As in EMHMC, the trajectories oscillate and the half-period of the longest oscillation, $T / 2 \approx 25$, defines the optimal integration time.
\label{fig:softAbsFunnelDiptych}}
\end{figure}

\begin{figure}[h!]
\centering
\includegraphics[width=7in]{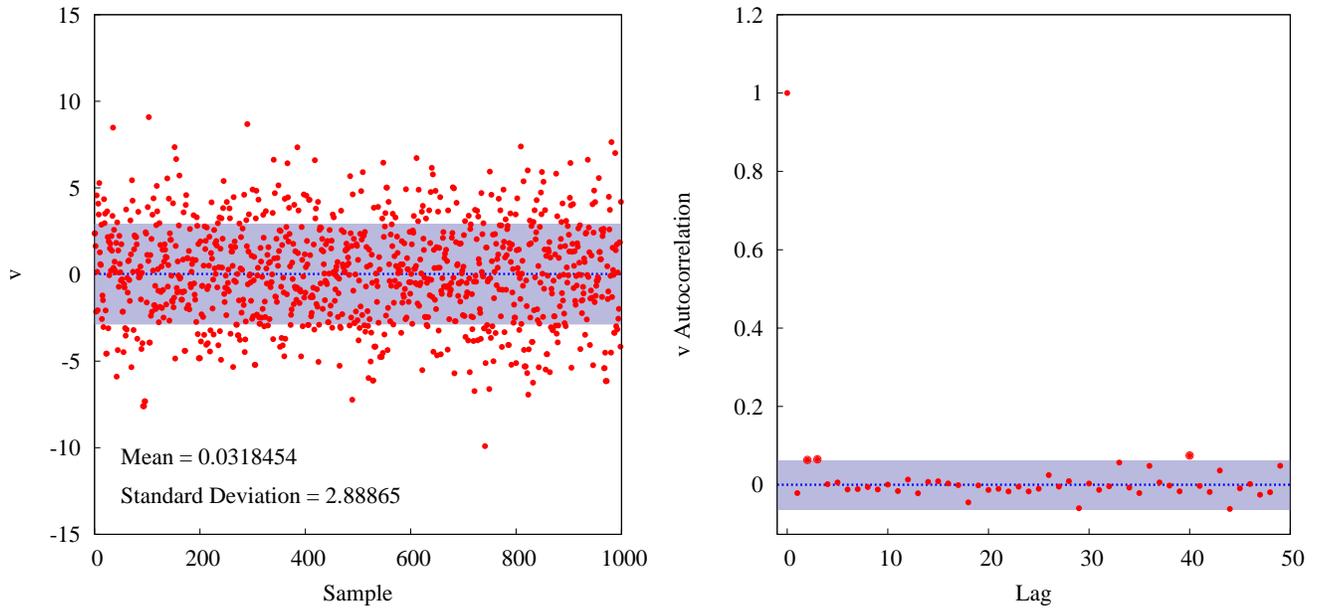}
\caption{The far-reaching trajectories of RMHMC dramatically reduce the autocorrelation of the samples which, despite the step-size adaptation, are consistent with the desired marginal $\mathcal{N} \!\left(0, 9\right)$.  Note that significantly fewer lags shown here than above.
\label{fig:softAbsFunnelTrace}}
\end{figure}

On the other hand, the SoftAbs metric, here with $\alpha = 10^{6}$, allows RMHMC to explore the entire distribution within a single trajectory (Figure \ref{fig:softAbsFunnelDiptych}).  Because the metric accounts for local curvature, the step-size can be adaptively tuned\footnote{The increased information encoded in the metric should admit a larger acceptance rate, $r$, for RMHMC than the EMHMC case of $r = 0.65$.  Motivated by some simple experiments, here the target rate for RMHMC is set to $r = 0.95$.} without introducing any bias.  The huge autocorrelations of EMHMC vanish (Figure \ref{fig:softAbsFunnelTrace}) and, despite the increased computation required for each transition, RMHMC yields a more efficient generation of effective samples (Table \ref{tab:benchmark}).  

Nominally, the $O \! \left( N^{3} \right)$ computational burden of RMHMC is significantly worse than the $O \! \left( N \right)$ burden of EMHMC.  The pathological behavior of distributions like the funnel, however, scales much faster, often exponentially, and the benefit of RMHMC with the SoftAbs metric only increases with dimension.  Moreover, this concern ignores the burden of computing the potential itself which, as in the case of Bayesian posteriors with many data, can overwhelm the $O \! \left( N^{3} \right)$ burden entirely.

\clearpage

\onecolumngrid

\clearpage

\clearpage

\clearpage

\twocolumngrid

\section{Conclusions and Future Work}

By smoothly regularizing the eigendecomposition of the Hessian, the SoftAbs metric admits a general implementation of RMHMC robust against the many pathologies to which EMHMC can be vulnerable.  Despite its apparently steep computational burden, the SoftAbs metric allows for practical inference on complex models never before deemed feasible.

Potential to reduce that burden might be found by looking deeper into Riemannian geometry.  Use of frame bundles \cite{Spivak2005b} to transport the Hessian eigenbasis along a trajectory, for example, could be more efficient than recomputing the eigendecomposition at each point.  

Moreover, further insight into the metric itself may be found in understanding the geometric consequences of the SoftAbs mapping.  What effect, for example, does the regularization of the SoftAbs mapping have on the geometric curvature of the manifold?  Perhaps this parameterized regularization be related to Ricci flow~\cite{Chow2004}, or other smoothing diffeomorphisms.

\section{Acknowledgements}

I warmly thank Bob Carpenter, Joe Formaggio, Mark Girolami, and Chris Jones for helpful comments and suggestions, as well as Andrew Gelman and the Stan team for their generous hospitality.

\vspace{2in}

\section*{Appendix A: Approximations to the SoftAbs Metric}
\label{sec:approximations}

\subsection{Approximations}

The SoftAbs metric requires an eigendecomposition and all third-order derivatives of the potential, both of which can be sufficiently burdensome as to render the metric unfeasible for very large problems.  Transforming various approximations to the Hessian with the SoftAbs map, however, yields a series of approximations that offer less intense alternatives that may be useful for certain distributions.

\subsubsection{Diagonal}

Ignoring the local decorrelation of the potential and instead focusing on the rescaling of the parameters, we can simply take
\begin{equation*}
\mathbf{H} \approx \mathrm{Diag} \left( H_{ii} \right),
\end{equation*}
with
\begin{equation*}
\sabs{H} \approx \mathrm{Diag} \left( H_{ii} \coth \alpha H_{ii} \right).
\end{equation*}

\begin{figure}[h!]
\centering
\includegraphics[width=7in]{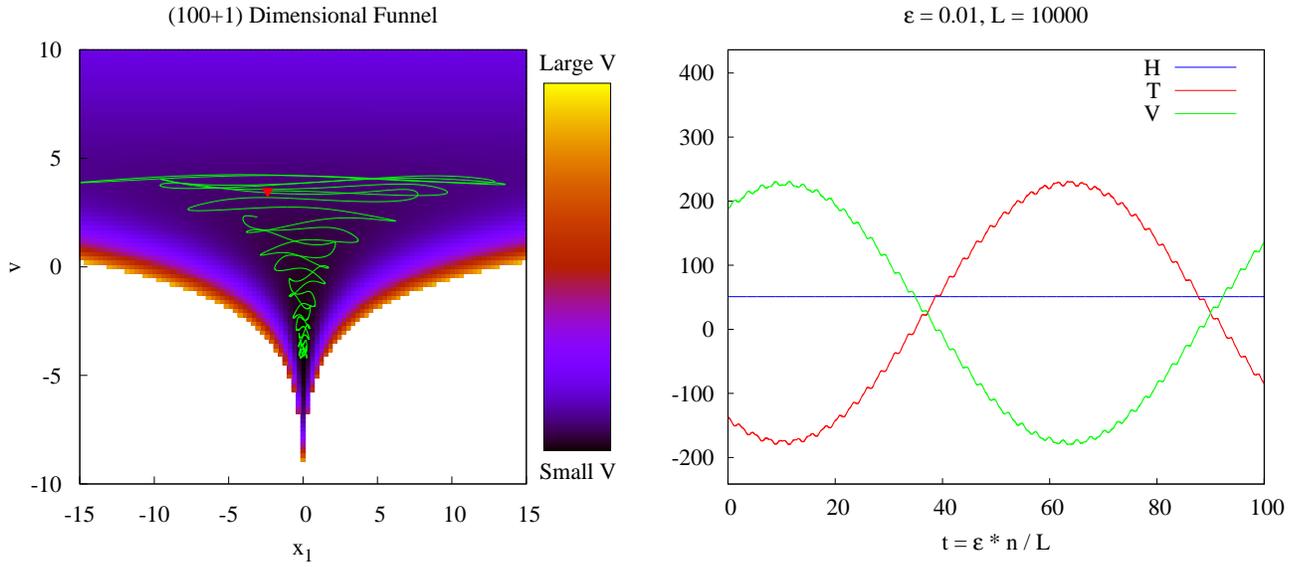}
\caption{Although clearly more ragged than the trajectories with the full SoftAbs metric, the trajectories with the diagonal SoftAbs metric explore almost the same expanse as the exact metric.
\label{fig:diagSoftAbsFunnelDiptych}}
\end{figure}

\begin{figure}[h!]
\centering
\includegraphics[width=7in]{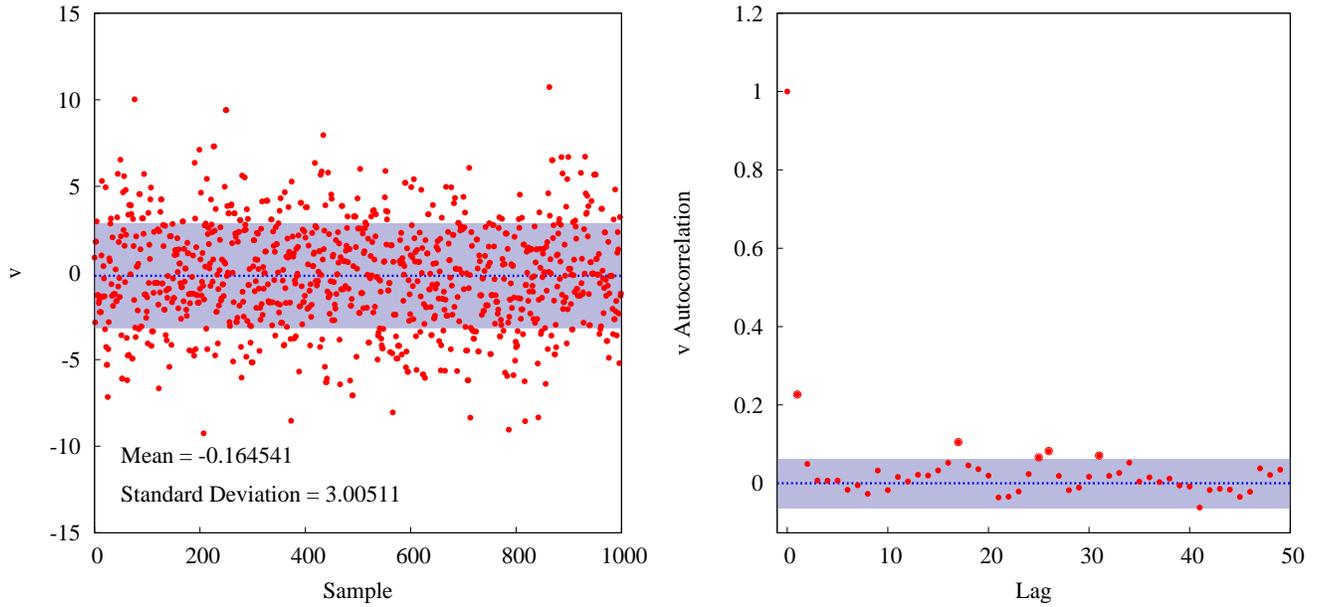}
\caption{The samples drawn with the diagonal SoftAbs metric are consistent with the true marginal for $v$, albeit at the expense of slightly higher autocorrelations.
\label{fig:diagSoftAbsFunnelTrace}}
\end{figure}

The computational burden of the resulting evolution scales as $O \! \left(N^{2} \right)$, much better than the $O \! \left(N^{3} \right)$ of the full SoftAbs metric, and, because the Hessian of the funnel is almost diagonal, the approximation loses little information.  Consequently, the diagonal approximation performs exceptionally well in this case (Table \ref{tab:approxBenchmark}, Figures \ref{fig:diagSoftAbsFunnelDiptych}, \ref{fig:diagSoftAbsFunnelTrace}).  Note that, on more correlated distributions the approximation will be more limiting and the performance will suffer.

\begin{widetext}

\begin{table*}[b!]
\begin{center}
	\footnotesize
	\renewcommand{\arraystretch}{1.2}
	\begin{tabular}{c@{}c@{}c@{}c@{}c@{}c@{}c@{}c}
	\rowcolor[gray]{0.8} \tabHead{Algorithm} & \tabHead{Warm-Up Iterations} & \tabHead{Samples} & \tabHead{$\mathbf{\epsilon}$} & \tabHead{Accept Rate} & \tabHead{CPU Time (s)} & \tabHead{ESS} & \tabHead{ESS / Time ($\mathbf{s^{1}}$)} \\
	 EMHMC & $10^{3}$ & $10^{5}$ & 0.001 & 0.999 & 1627 & 70.3 & 0.0432 \\
	 \rowcolor[gray]{0.8} RMHMC (SoftAbs) & $10^{3}$ & $10^{3}$ & 0.21 & 0.946 & 6282 & 856 & 0.136 \\
	 RMHMC (Diag SoftAbs) & $10^{3}$ & $10^{3}$ & 0.49 & 0.805 & 7.694 & 633 & 82.3 \\
	\end{tabular}
	\caption{The diagonal approximation to the SoftAbs metric dramatically out-performs the full SoftAbs metric when applied to the funnel distribution, because the Hessian of the funnel potential is almost diagonal.  As before, the step-size of the diagonal SoftAbs chain was determined with dual-averaging but with a target acceptance rate of $r = 0.8$.}
	\label{tab:approxBenchmark}
\end{center}
\end{table*}

\end{widetext}

\clearpage
\onecolumngrid

\clearpage

\begin{figure}[h!]
\centering
\includegraphics[width=7in]{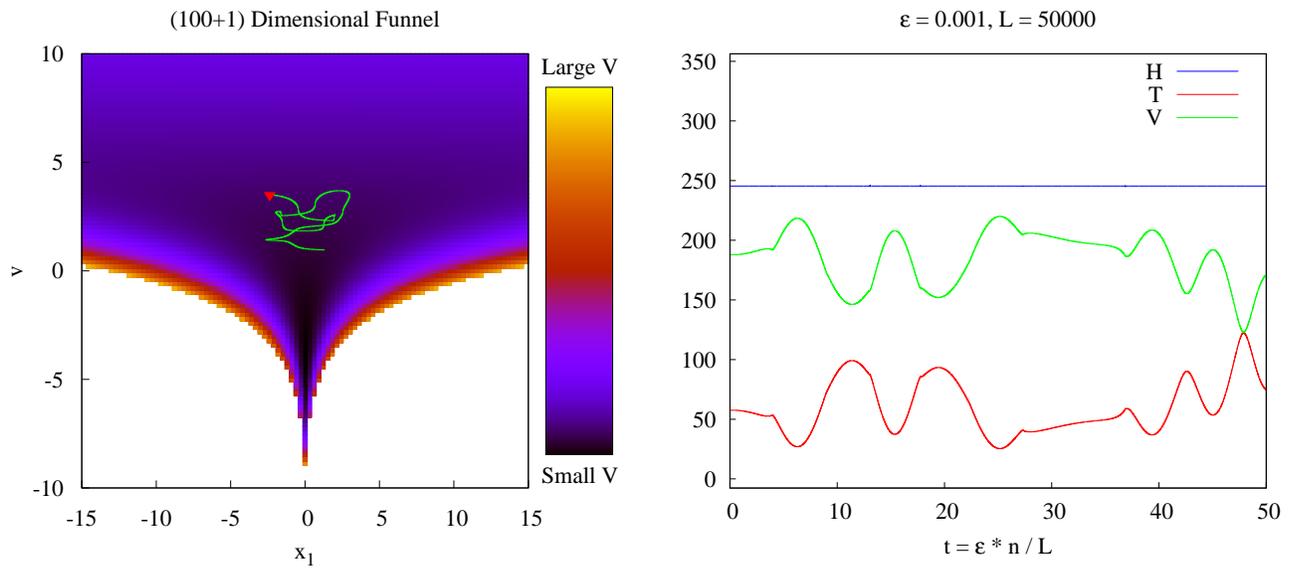}
\caption{When $\alpha$ has to be kept small to ensure the stability of the numerical evolution, the benefit of the dynamic metric is substantially reduced.  The oscillating trajectories become much more complicated and, as in the EMHMC case, the variation of the potential over each trajectory is limited.
\label{fig:diagOuterSoftAbsFunnelDiptych}}
\end{figure}

\begin{figure}[h!]
\centering
\includegraphics[width=7in]{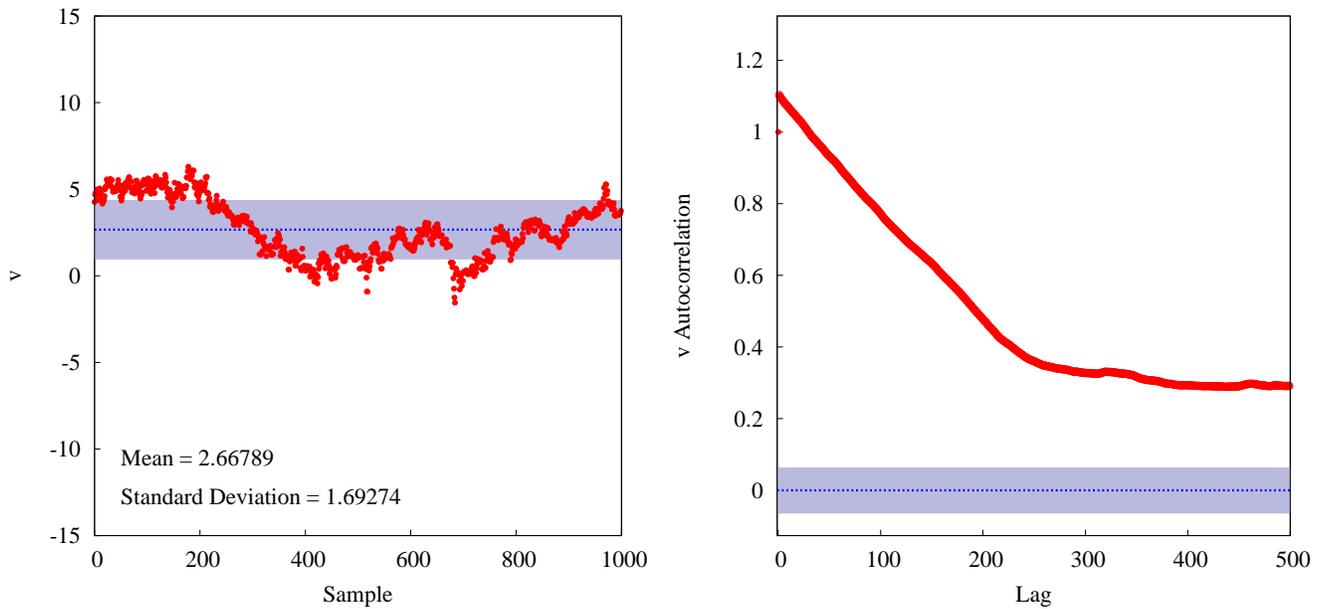}
\caption{The limited variation in potential induces random walk behavior in the diagonal outer-product approximation, while the small step-sizes required for stable evolution dramatically reduce the efficiency of each transition.  Moreover, dual-averaging adaptation of the step-size introduces the same bias seen in adaptive tuning of EMHMC.
\label{fig:diagOuterSoftAbsFunnelTrace}}
\end{figure}

\clearpage

\twocolumngrid

\subsubsection{Outer-Product}

A common approximation to the Hessian is the outer-product of the gradient~\cite{Bishop2007},
\begin{equation*}
\mathbf{H} \approx \mathbf{g} \mathbf{g}^{T},
\end{equation*}
where $\mathbf{g} = \partial V$.  Propagating this approximation through the SoftAbs map gives
\begin{equation*}
\sabs{H} \approx \frac{ \left( \mathbf{g} \cdot \mathbf{g} \right) }{ \sinh \left( \alpha \, \mathbf{g} \cdot \mathbf{g} \right) } \left( \mathbb{I} + \frac{ \cosh \left( \alpha \, \mathbf{g} \cdot \mathbf{g} \right) - 1 }{ \left( \mathbf{g} \cdot \mathbf{g} \right) } \mathbf{g} \mathbf{g}^{T} \right).
\end{equation*}
Note that this is equivalent to the ``background-score'' metric proposed in~\cite{Betan2011} with functions scaling the rank-one update as well as the entire metric.  

Evolution with this metric requires no derivatives beyond the Hessian, and with a careful implementation the computation scales as $O \! \left( N^{2} \right)$\footnote{In fact, with automatic differentiation techniques that can compute Hessian-vector products in linear time~\cite{Bishop2007}, the computation can be reduced further to $O \! \left( N \right)$.}.  The global coefficient, however, renders the metric almost singular in all but the simplest problems.  

Along typical trajectories the norm of the gradient is far from zero, and the $\sinh$ becomes highly nonlinear.  Consequently, the numerical evolution becomes unstable unless $\alpha \ll 1$ or $\epsilon \ll 1$, at which point the evolution becomes impractical.  For example, with $\alpha = 1$ the step-size must be smaller than floating-point precision before the Metropolis accept rate rises above 0.5.

\subsubsection{Diagonal Outer-Product}

The severe non-linearities of the outer-product approximation can be avoided by considering only the diagonal elements of the outer-product,
\begin{equation*}
\mathbf{H} \approx \mathrm{Diag} \left( g_{i}^{2} \right),
\end{equation*}
which gives the metric
\begin{equation*}
\sabs{H} \approx \mathrm{Diag} \left( g_{i}^{2} \coth \alpha g_{i}^{2} \right).
\end{equation*}
As above, $g_{i} = \partial_{i} V$.

Although this approximation avoids the highly-nonlinear coefficient above, the typically large values of the $g_{i}^{2}$ do induce large gradients in the Hamiltonian evolution.  Taking $\alpha = 1$ yields stable trajectories, but the strong regularization reduces the log determinant's ability to increase the variation in the potential (Figure \ref{fig:diagOuterSoftAbsFunnelDiptych}) and limits the adaptability of the metric.  The limited adaptability makes the Markov chain vulnerable to bias when adapting the integrator step-size and, consequently the samples exhibit the same random walk behavior and bias towards large $v$ seen in the EMHMC case (Figure \ref{fig:diagOuterSoftAbsFunnelTrace}).

\section*{Appendix B: A Symplectic Integrator For RMHMC}
\label{sec:integrator}

One of the challenges of RMHMC is that the non-separable Hamiltonian requires a more sophisticated, and costly, symplectic integrator than EMHMC~\cite{Leimkuhler2004, Hairer2006}.  In order to construct such an integrator first rewrite the Hamiltonian as
\begin{equation*}
H = \tau + \phi,
\end{equation*}
where
\begin{equation*}
\tau = \frac{1}{2} \mathbf{p}^{T} \cdot \mathbf{\Sigma}^{-1} \! \left( \mathbf{q} \right) \cdot \mathbf{p}
\end{equation*}
and
\begin{equation*}
\phi = \frac{1}{2} \log | \mathbf{\Sigma} \! \left( \mathbf{q} \right) | + V.
\end{equation*}

Both proper scalar functions,  $\tau$ and $\phi$ motivate a natural splitting of the Hamiltonian which yields the integrator
\begin{equation} \label{splitting}
\Phi_{t} = \hat{\phi}_{\frac{t}{2}} \circ \hat{\tau}_{\frac{t}{2}} \circ \hat{T}_{t} \circ \hat{\tau}_{\frac{t}{2}} \circ \hat{\phi}_{\frac{t}{2}},
\end{equation}
with
\begin{align*}
\hat{\phi} &= \frac{ \partial \phi}{ \partial q^{i} } \frac{ \partial }{ \partial p_{i}} \\
\hat{\tau} &= \frac{ \partial \tau}{ \partial q^{i} } \frac{ \partial }{ \partial p_{i} } \\
\hat{T} &= \frac{ \partial \tau}{ \partial p^{i} } \frac{ \partial }{ \partial q_{i} }.
\end{align*}

Of the individual operators, only $\hat{\phi}$ is separable and analytically calculable.  The non-separable operators, $\hat{\tau}_{\frac{t}{2}} \circ \hat{T}_{t} \circ \hat{\tau}_{\frac{t}{2}}$, require the implicit generalized leapfrog algorithm (Algorithm \ref{algo:evolution}).  Efficient gradient computations complete the implementation (Algorithm \ref{algo:gradients}).

\begin{algorithm}
\caption{The spitting \eqref{splitting} admits a discrete integration of Hamilton's equations, approximating the evolution of the current position, $\mathbf{q}$, and momentum, $\mathbf{p}$, for a time, $\Delta t = \epsilon$.  Because they are implicit, the first two steps of the generalized leapfrog must be solved with fixed-point iterations; in order to avoid position-dependent divergences causing bias in the Markov transitions, the iterations are continued until a given threshold instead of a set number of iterations. }
\label{algo:evolution}
\begin{algorithmic}

\STATE $\mathbf{p} \gets \mathbf{p} - \frac{\epsilon}{2} \partial \phi / \partial \mathbf{q} \left( \mathbf{p}, \mathbf{q} \right) $ \hfill {\color{comment} $\hat{\phi}$ }
\STATE 
\STATE $ \mbox{\boldmath{$\rho$}} \gets \mathbf{p}$ \hfill {\color{comment} $\hat{\tau}_{\frac{t}{2}} \circ \hat{T}_{t} \circ \hat{\tau}_{\frac{t}{2}}$ }
\WHILE{$\Delta p > \delta $}
   \STATE $\mathbf{p}' \gets \mbox{\boldmath{$\rho$}} - \frac{\epsilon}{2} \cdot \partial \tau  / \partial \mathbf{q} \left( \mathbf{p}, \mathbf{q} \right) $
   \STATE $\Delta p = \max_{i} \left\{ \left| p_{i} - p_{i}' \right| \right\} $
   \STATE $\mathbf{p} \gets \mathbf{p}'$
\ENDWHILE
\STATE
\STATE $ \mbox{\boldmath{$\sigma$}} \gets \mathbf{q}$
\WHILE{$\Delta q > \delta $}
    \STATE $\mathbf{q}' \gets \mbox{\boldmath{$\sigma$}} + \frac{\epsilon}{2} \cdot \partial \tau / \partial \mathbf{p} \left( \mathbf{p}, \mbox{\boldmath{$\sigma$}} \right) + \frac{\epsilon}{2} \cdot \partial \tau / \partial \mathbf{p} \left( \mathbf{p}, \mathbf{q} \right) $
    \STATE $\Delta q = \max_{i} \left\{ \left| q_{i} - q_{i}' \right| \right\}$
    \STATE $\mathbf{q} \gets \mathbf{q}'$
\ENDWHILE
\STATE
\STATE $\mathbf{p} \gets \mathbf{p} - \frac{\epsilon}{2} \cdot \partial \tau / \partial \mathbf{q} \left( \mathbf{p}, \mathbf{q} \right) $
\STATE
\STATE $\mathbf{p} \gets \mathbf{p} - \frac{\epsilon}{2} \partial \phi / \partial \mathbf{q} \left( \mathbf{p}, \mathbf{q} \right) $ \hfill {\color{comment} $\hat{\phi}$ }

\end{algorithmic}
\end{algorithm}

\begin{algorithm}
\caption{With careful manipulation and caching of intermediate matrices, the gradients necessary for Hamiltonian evolution (Algorithm \ref{algo:evolution}) can be computed in $O \! \left( N^{3} \right)$.  Note that each trace can be computed in $O \! \left( N^{2} \right)$ as $\mathrm{Tr} \left[ \mathbf{A} \cdot \mathbf{B} \right] = \sum_{i, j} A_{ij} B_{ji}$.}
\label{algo:gradients}
\begin{algorithmic}

\REQUIRE $\alpha$ \hfill {\color{comment} \textit{Regularization parameter} }
\REQUIRE $\mathbf{H}$ \hfill {\color{comment} \textit{$ \partial^{2} V / \partial q^{i} \partial q^{j}  $} }
\REQUIRE $\partial \mathbf{H}$ \hfill {\color{comment} \textit{$ \partial^{3} V / \partial q^{n} \partial q^{i} \partial q^{j} $} }
\REQUIRE $\mbox{\boldmath{$\lambda$}}, \mathbf{Q}$ \hfill {\color{comment} \textit{Eigendecomposition of $\mathbf{H}$} }
\STATE
\STATE \textbf{function} $\partial \tau  / \partial \mathbf{p} \left( \mathbf{p}, \mathbf{q} \right)$
\STATE $\quad \mathbf{return} \;\; \mathbf{Q} \cdot \sabsGreek{\lambda}^{-1} \cdot \mathbf{Q}^{T} \cdot \mathbf{p}$
\STATE
\STATE \textbf{function} $\partial \tau  / \partial \mathbf{q} \left( \mathbf{p}, \mathbf{q} \right)$
\STATE $\quad J_{ij} \gets \dfrac{ \lambda_{i} \coth \alpha \lambda_{i} - \lambda_{j} \coth \alpha \lambda_{j} }{ \lambda_{i} - \lambda_{j} } \left( 1 - \delta_{ij} \right) $
\STATE $\quad \quad \quad \quad + \dfrac{ \partial }{ \partial \lambda_{i} } \lambda_{i} \coth \alpha \lambda_{i} \delta_{ij}$
\STATE $\quad \mathbf{D} \gets \mathrm{Diag} \left( \left( \mathbf{Q}^{T} \cdot \mathbf{p} \right)_{i} \right)$
\STATE $\quad \mathbf{M} \gets \mathbf{Q} \cdot \mathbf{D} \cdot \mathbf{J} \cdot \mathbf{D} \cdot \mathbf{Q}^{T}$
\STATE $\quad \mathbf{for} \; n = 1 \; \mathrm{to} \; N \; \mathbf{do}$
\STATE $\quad\quad \delta_{n} \gets \frac{1}{2} \mathrm{Tr} \left[ - \mathbf{M} \cdot \partial_{n} \mathbf{H} \right]$
\STATE $\quad \mathbf{end \; for}$
\STATE $\quad \mathbf{return} \;\; \mbox{\boldmath{$\delta$}}$

\STATE
\STATE \textbf{function} $\partial \phi  / \partial \mathbf{q} \left( \mathbf{p}, \mathbf{q} \right)$
\STATE $\quad J_{ij} \gets \dfrac{ \lambda_{i} \coth \alpha \lambda_{i} - \lambda_{j} \coth \alpha \lambda_{j} }{ \lambda_{i} - \lambda_{j} } \left( 1 - \delta_{ij} \right) $
\STATE $\quad \quad \quad \quad + \dfrac{ \partial }{ \partial \lambda_{i} } \lambda_{i} \coth \alpha \lambda_{i} \delta_{ij}$
\STATE $\quad \mathbf{R} \gets \mathrm{Diag} \left( \dfrac{1}{\lambda_{i} \coth \alpha \lambda_{i}} \right)$
\STATE $\quad \mathbf{M} \gets \mathbf{Q} \left( \mathbf{R} \circ \mathbf{J} \right) \mathbf{Q}^{T}$
\STATE $\quad \mathbf{for} \; n = 1 \; \mathrm{to} \; N \; \mathbf{do}$
\STATE $\quad\quad \delta_{n} \gets \frac{1}{2} \mathrm{Tr} \left[ \mathbf{M} \cdot \partial_{n} \mathbf{H} \right] + \partial_{n} V$
\STATE $\quad \mathbf{end \; for}$
\STATE $\quad \mathbf{return} \;\; \mbox{\boldmath{$\delta$}}$

\end{algorithmic}
\end{algorithm}

\clearpage

\bibliography{softAbsMetric}

\begin{thebibliography}{21}%
\makeatletter
\providecommand \@ifxundefined [1]{%
 \@ifx{#1\undefined}
}%
\providecommand \@ifnum [1]{%
 \ifnum #1\expandafter \@firstoftwo
 \else \expandafter \@secondoftwo
 \fi
}%
\providecommand \@ifx [1]{%
 \ifx #1\expandafter \@firstoftwo
 \else \expandafter \@secondoftwo
 \fi
}%
\providecommand \natexlab [1]{#1}%
\providecommand \enquote  [1]{``#1''}%
\providecommand \bibnamefont  [1]{#1}%
\providecommand \bibfnamefont [1]{#1}%
\providecommand \citenamefont [1]{#1}%
\providecommand \href@noop [0]{\@secondoftwo}%
\providecommand \href [0]{\begingroup \@sanitize@url \@href}%
\providecommand \@href[1]{\@@startlink{#1}\@@href}%
\providecommand \@@href[1]{\endgroup#1\@@endlink}%
\providecommand \@sanitize@url [0]{\catcode `\\12\catcode `\$12\catcode
  `\&12\catcode `\#12\catcode `\^12\catcode `\_12\catcode `\%12\relax}%
\providecommand \@@startlink[1]{}%
\providecommand \@@endlink[0]{}%
\providecommand \url  [0]{\begingroup\@sanitize@url \@url }%
\providecommand \@url [1]{\endgroup\@href {#1}{\urlprefix }}%
\providecommand \urlprefix  [0]{URL }%
\providecommand \Eprint [0]{\href }%
\@ifxundefined \urlstyle {%
  \providecommand \doi  [0]{\begingroup \@sanitize@url \@doi}%
  \providecommand \@doi [1]{\endgroup \@@startlink {\doibase
  #1}doi:\discretionary {}{}{}#1\@@endlink }%
}{%
  \providecommand \doi  [0]{doi:\discretionary{}{}{}\begingroup
  \urlstyle{rm}\Url }%
}%
\providecommand \doibase [0]{http://dx.doi.org/}%
\providecommand \Doi [0]{\begingroup \@sanitize@url \@Doi }%
\providecommand \@Doi  [1]{\endgroup\@@startlink{\doibase#1}\@@Doi}%
\providecommand \@@Doi [1]{#1\@@endlink}%
\providecommand \selectlanguage [0]{\@gobble}%
\providecommand \bibinfo  [0]{\@secondoftwo}%
\providecommand \bibfield  [0]{\@secondoftwo}%
\providecommand \translation [1]{[#1]}%
\providecommand \BibitemOpen [0]{}%
\providecommand \bibitemStop [0]{}%
\providecommand \bibitemNoStop [0]{.\EOS\space}%
\providecommand \EOS [0]{\spacefactor3000\relax}%
\providecommand \BibitemShut  [1]{\csname bibitem#1\endcsname}%
\bibitem [{\citenamefont {Betancourt}\ and\ \citenamefont
  {Stein}(2011)}]{Betan2011}%
  \BibitemOpen
  \bibfield  {author} {\bibinfo {author} {\bibfnamefont {M.}~\bibnamefont
  {Betancourt}}\ and\ \bibinfo {author} {\bibfnamefont {L.~C.}\ \bibnamefont
  {Stein}},\ }\href@noop {} { (\bibinfo {year} {2011})},\ \Eprint
  {http://arxiv.org/abs/1112.4118} {arXiv:1112.4118 [stat]} \BibitemShut
  {NoStop}%
\bibitem [{\citenamefont {Duane}\ \emph {et~al.}(1987)\citenamefont {Duane},
  \citenamefont {Kennedy}, \citenamefont {Pendleton},\ and\ \citenamefont
  {Roweth}}]{Duane1987}%
  \BibitemOpen
  \bibfield  {author} {\bibinfo {author} {\bibfnamefont {S.}~\bibnamefont
  {Duane}}, \bibinfo {author} {\bibfnamefont {A.}~\bibnamefont {Kennedy}},
  \bibinfo {author} {\bibfnamefont {B.~J.}\ \bibnamefont {Pendleton}}, \ and\
  \bibinfo {author} {\bibfnamefont {D.}~\bibnamefont {Roweth}},\ }\Doi
  {10.1016/0370-2693(87)91197-X} {\bibfield  {journal} {\bibinfo  {journal}
  {Physics Letters B},\ }\textbf {\bibinfo {volume} {195}},\ \bibinfo {pages}
  {216 } (\bibinfo {year} {1987})},\ ISSN \bibinfo {issn}
  {0370-2693}\BibitemShut {NoStop}%
\bibitem [{\citenamefont {Neal}(2011)}]{Neal2011}%
  \BibitemOpen
  \bibfield  {author} {\bibinfo {author} {\bibfnamefont {R.}~\bibnamefont
  {Neal}},\ }in\ \href
  {http://www.cs.utoronto.ca/~radford/ham-mcmc.abstract.html} {\emph {\bibinfo
  {booktitle} {Handbook of Markov Chain Monte Carlo}}},\ \bibinfo {editor}
  {edited by\ \bibinfo {editor} {\bibfnamefont {S.}~\bibnamefont {Brooks}},
  \bibinfo {editor} {\bibfnamefont {A.}~\bibnamefont {Gelman}}, \bibinfo
  {editor} {\bibfnamefont {G.~L.}\ \bibnamefont {Jones}}, \ and\ \bibinfo
  {editor} {\bibfnamefont {X.-L.}\ \bibnamefont {Meng}}}\ (\bibinfo
  {publisher} {CRC Press},\ \bibinfo {address} {New York},\ \bibinfo {year}
  {2011})\BibitemShut {NoStop}%
\bibitem [{\citenamefont {Girolami}\ and\ \citenamefont
  {Calderhead}(2011)}]{Girolami2011}%
  \BibitemOpen
  \bibfield  {author} {\bibinfo {author} {\bibfnamefont {M.}~\bibnamefont
  {Girolami}}\ and\ \bibinfo {author} {\bibfnamefont {B.}~\bibnamefont
  {Calderhead}},\ }\Doi {10.1111/j.1467-9868.2010.00765.x} {\bibfield
  {journal} {\bibinfo  {journal} {Journal of the Royal Statistical Society:
  Series B (Statistical Methodology)},\ }\textbf {\bibinfo {volume} {73}},\
  \bibinfo {pages} {123} (\bibinfo {year} {2011})},\ ISSN \bibinfo {issn}
  {1467-9868}\BibitemShut {NoStop}%
\bibitem [{\citenamefont {Amari}\ and\ \citenamefont
  {Nagaoka}(2007)}]{Amari2007}%
  \BibitemOpen
  \bibfield  {author} {\bibinfo {author} {\bibfnamefont {S.}~\bibnamefont
  {Amari}}\ and\ \bibinfo {author} {\bibfnamefont {H.}~\bibnamefont
  {Nagaoka}},\ }\href@noop {} {\emph {\bibinfo {title} {Methods of information
  geometry}}},\ Vol.\ \bibinfo {volume} {191}\ (\bibinfo  {publisher} {Amer
  Mathematical Society},\ \bibinfo {year} {2007})\BibitemShut {NoStop}%
\bibitem [{\citenamefont {Spivak}(2005){\natexlab{a}}}]{Spivak2005a}%
  \BibitemOpen
  \bibfield  {author} {\bibinfo {author} {\bibfnamefont {M.}~\bibnamefont
  {Spivak}},\ }\href@noop {} {\emph {\bibinfo {title} {A Comprehensive
  Introduction to Differential Geometry}}},\ Vol.~\bibinfo {volume} {1}\
  (\bibinfo  {publisher} {Publish or Perish, Inc},\ \bibinfo {address}
  {Houston, Texas},\ \bibinfo {year} {2005})\BibitemShut {NoStop}%
\bibitem [{\citenamefont {Celis}\ \emph {et~al.}(1985)\citenamefont {Celis},
  \citenamefont {Dennis},\ and\ \citenamefont {A.}}]{Celis1985}%
  \BibitemOpen
  \bibfield  {author} {\bibinfo {author} {\bibfnamefont {M.}~\bibnamefont
  {Celis}}, \bibinfo {author} {\bibfnamefont {J.~E.}\ \bibnamefont {Dennis}}, \
  and\ \bibinfo {author} {\bibfnamefont {T.~R.}\ \bibnamefont {A.}},\ }in\
  \href@noop {} {\emph {\bibinfo {booktitle} {Numerical Optimization 1984}}},\
  \bibinfo {editor} {edited by\ \bibinfo {editor} {\bibfnamefont
  {P.}~\bibnamefont {Boggs}}, \bibinfo {editor} {\bibfnamefont
  {R.}~\bibnamefont {Byrd}}, \ and\ \bibinfo {editor} {\bibfnamefont
  {R.}~\bibnamefont {Schnabel}}}\ (\bibinfo  {publisher} {SIAM},\ \bibinfo
  {address} {Philadelphia},\ \bibinfo {year} {1985})\BibitemShut {NoStop}%
\bibitem [{\citenamefont {Moler}\ and\ \citenamefont
  {Van~Loan}(2003)}]{Moler2003}%
  \BibitemOpen
  \bibfield  {author} {\bibinfo {author} {\bibfnamefont {C.}~\bibnamefont
  {Moler}}\ and\ \bibinfo {author} {\bibfnamefont {C.}~\bibnamefont
  {Van~Loan}},\ }\href@noop {} {\bibfield  {journal} {\bibinfo  {journal} {SIAM
  review},\ }\textbf {\bibinfo {volume} {45}},\ \bibinfo {pages} {3} (\bibinfo
  {year} {2003})}\BibitemShut {NoStop}%
\bibitem [{\citenamefont {Aizu}(1963)}]{Aizu1963}%
  \BibitemOpen
  \bibfield  {author} {\bibinfo {author} {\bibfnamefont {K.}~\bibnamefont
  {Aizu}},\ }\Doi {10.1063/1.1724318} {\bibfield  {journal} {\bibinfo
  {journal} {Journal of Mathematical Physics},\ }\textbf {\bibinfo {volume}
  {4}},\ \bibinfo {pages} {762} (\bibinfo {year} {1963})}\BibitemShut {NoStop}%
\bibitem [{\citenamefont {Wilcox}(1967)}]{Wilcox1967}%
  \BibitemOpen
  \bibfield  {author} {\bibinfo {author} {\bibfnamefont {R.~M.}\ \bibnamefont
  {Wilcox}},\ }\Doi {10.1063/1.1705306} {\bibfield  {journal} {\bibinfo
  {journal} {Journal of Mathematical Physics},\ }\textbf {\bibinfo {volume}
  {8}},\ \bibinfo {pages} {962} (\bibinfo {year} {1967})}\BibitemShut {NoStop}%
\bibitem [{\citenamefont {Magnus}\ and\ \citenamefont
  {Neudecker}(2007)}]{Magnus2007}%
  \BibitemOpen
  \bibfield  {author} {\bibinfo {author} {\bibfnamefont {J.}~\bibnamefont
  {Magnus}}\ and\ \bibinfo {author} {\bibfnamefont {H.}~\bibnamefont
  {Neudecker}},\ }\href@noop {} {\emph {\bibinfo {title} {Matrix Differential
  Calculus with Applications in Statistics and Econometrics}}}\ (\bibinfo
  {publisher} {Wiley},\ \bibinfo {address} {New York},\ \bibinfo {year}
  {2007})\BibitemShut {NoStop}%
\bibitem [{\citenamefont {Neal}(2003)}]{Neal2003}%
  \BibitemOpen
  \bibfield  {author} {\bibinfo {author} {\bibfnamefont {R.}~\bibnamefont
  {Neal}},\ }\href@noop {} {\bibfield  {journal} {\bibinfo  {journal} {Annals
  of statistics},\ \bibinfo {pages} {705}} (\bibinfo {year}
  {2003})}\BibitemShut {NoStop}%
\bibitem [{\citenamefont {Gelman}\ \emph {et~al.}(2004)\citenamefont {Gelman},
  \citenamefont {Carlin}, \citenamefont {Stern},\ and\ \citenamefont
  {Rubin}}]{Gelman2004}%
  \BibitemOpen
  \bibfield  {author} {\bibinfo {author} {\bibfnamefont {A.}~\bibnamefont
  {Gelman}}, \bibinfo {author} {\bibfnamefont {J.}~\bibnamefont {Carlin}},
  \bibinfo {author} {\bibfnamefont {H.}~\bibnamefont {Stern}}, \ and\ \bibinfo
  {author} {\bibfnamefont {D.}~\bibnamefont {Rubin}},\ }\href@noop {} {\emph
  {\bibinfo {title} {Bayesian Data Analysis}}}\ (\bibinfo  {publisher} {Chapman
  \& Hall/CRC Press},\ \bibinfo {address} {Boca Raton, Florida},\ \bibinfo
  {year} {2004})\BibitemShut {NoStop}%
\bibitem [{\citenamefont {{Murray}}\ and\ \citenamefont {{Prescott
  Adams}}(2010)}]{Murray2010}%
  \BibitemOpen
  \bibfield  {author} {\bibinfo {author} {\bibfnamefont {I.}~\bibnamefont
  {{Murray}}}\ and\ \bibinfo {author} {\bibfnamefont {R.}~\bibnamefont
  {{Prescott Adams}}},\ }\href@noop {} {\bibfield  {journal} {\bibinfo
  {journal} {ArXiv e-prints}} (\bibinfo {year} {2010})},\ \Eprint
  {http://arxiv.org/abs/1006.0868} {arXiv:1006.0868 [stat.CO]} \BibitemShut
  {NoStop}%
\bibitem [{\citenamefont {{Hoffman}}\ and\ \citenamefont
  {{Gelman}}(2011)}]{Hoffman2011}%
  \BibitemOpen
  \bibfield  {author} {\bibinfo {author} {\bibfnamefont {M.~D.}\ \bibnamefont
  {{Hoffman}}}\ and\ \bibinfo {author} {\bibfnamefont {A.}~\bibnamefont
  {{Gelman}}},\ }\href@noop {} {\bibfield  {journal} {\bibinfo  {journal}
  {ArXiv e-prints}} (\bibinfo {year} {2011})},\ \Eprint
  {http://arxiv.org/abs/1111.4246} {arXiv:1111.4246 [stat.CO]} \BibitemShut
  {NoStop}%
\bibitem [{\citenamefont {Geyer}(2011)}]{Geyer2011}%
  \BibitemOpen
  \bibfield  {author} {\bibinfo {author} {\bibfnamefont {C.}~\bibnamefont
  {Geyer}},\ }in\ \href {http://www.mcmchandbook.net/HandbookChapter1.pdf}
  {\emph {\bibinfo {booktitle} {Handbook of Markov Chain Monte Carlo}}},\
  \bibinfo {editor} {edited by\ \bibinfo {editor} {\bibfnamefont
  {S.}~\bibnamefont {Brooks}}, \bibinfo {editor} {\bibfnamefont
  {A.}~\bibnamefont {Gelman}}, \bibinfo {editor} {\bibfnamefont {G.~L.}\
  \bibnamefont {Jones}}, \ and\ \bibinfo {editor} {\bibfnamefont {X.-L.}\
  \bibnamefont {Meng}}}\ (\bibinfo  {publisher} {CRC Press},\ \bibinfo
  {address} {New York},\ \bibinfo {year} {2011})\BibitemShut {NoStop}%
\bibitem [{\citenamefont {Spivak}(2005){\natexlab{b}}}]{Spivak2005b}%
  \BibitemOpen
  \bibfield  {author} {\bibinfo {author} {\bibfnamefont {M.}~\bibnamefont
  {Spivak}},\ }\href@noop {} {\emph {\bibinfo {title} {A Comprehensive
  Introduction to Differential Geometry}}},\ Vol.~\bibinfo {volume} {2}\
  (\bibinfo  {publisher} {Publish or Perish, Inc},\ \bibinfo {address}
  {Houston, Texas},\ \bibinfo {year} {2005})\BibitemShut {NoStop}%
\bibitem [{\citenamefont {Chow}\ and\ \citenamefont {Knopf}(2004)}]{Chow2004}%
  \BibitemOpen
  \bibfield  {author} {\bibinfo {author} {\bibfnamefont {B.}~\bibnamefont
  {Chow}}\ and\ \bibinfo {author} {\bibfnamefont {D.}~\bibnamefont {Knopf}},\
  }in\ \href@noop {} {\emph {\bibinfo {booktitle} {Mathematical Surveys and
  Monographs 110}}}\ (\bibinfo  {publisher} {American Mathematical Society},\
  \bibinfo {address} {Providence},\ \bibinfo {year} {2004})\BibitemShut
  {NoStop}%
\bibitem [{\citenamefont {Bishop}(2007)}]{Bishop2007}%
  \BibitemOpen
  \bibfield  {author} {\bibinfo {author} {\bibfnamefont {C.}~\bibnamefont
  {Bishop}},\ }\href@noop {} {\emph {\bibinfo {title} {Pattern Classification
  and Machine Learning}}}\ (\bibinfo  {publisher} {Springer},\ \bibinfo
  {address} {New York},\ \bibinfo {year} {2007})\BibitemShut {NoStop}%
\bibitem [{\citenamefont {Leimkuhler}\ and\ \citenamefont
  {Reich}(2004)}]{Leimkuhler2004}%
  \BibitemOpen
  \bibfield  {author} {\bibinfo {author} {\bibfnamefont {B.}~\bibnamefont
  {Leimkuhler}}\ and\ \bibinfo {author} {\bibfnamefont {S.}~\bibnamefont
  {Reich}},\ }\href@noop {} {\emph {\bibinfo {title} {Simulating Hamiltonian
  Dynamics}}}\ (\bibinfo  {publisher} {Cambridge University Press},\ \bibinfo
  {address} {New York},\ \bibinfo {year} {2004})\BibitemShut {NoStop}%
\bibitem [{\citenamefont {Hairer}\ \emph {et~al.}(2006)\citenamefont {Hairer},
  \citenamefont {Lubich},\ and\ \citenamefont {Wanner}}]{Hairer2006}%
  \BibitemOpen
  \bibfield  {author} {\bibinfo {author} {\bibfnamefont {E.}~\bibnamefont
  {Hairer}}, \bibinfo {author} {\bibfnamefont {C.}~\bibnamefont {Lubich}}, \
  and\ \bibinfo {author} {\bibfnamefont {G.}~\bibnamefont {Wanner}},\
  }\href@noop {} {\emph {\bibinfo {title} {Geometric Numerical Integration}}}\
  (\bibinfo  {publisher} {Springer},\ \bibinfo {address} {New York},\ \bibinfo
  {year} {2006})\BibitemShut {NoStop}%
\end{thebibliography}%

\end{document}